\documentclass{article}

\usepackage{amsmath,amssymb,amsthm,units,stmaryrd,upgreek}
\usepackage[usenames,dvipsnames]{xcolor}
\usepackage{yfonts}
\usepackage{units,stmaryrd}
\usepackage{color,xspace}
\usepackage{graphicx}
\usepackage[all]{xy}

\newtheorem{definition}{Definition}[section]
\newtheorem{lemm}{Lemma}[section]
\newtheorem{prop}{Proposition}[section]

\newtheorem{corollary}{Corollary}[section]

\newtheorem{theorem}{Theorem}[section]

\usepackage{bm}

\newcommand{\protname}{shifted projection}
\newcommand\agP{{\mathcal X}}
\newcommand\agQ{{\mathcal Y}}
\newcommand\prob{{\mathbb P}}
\newcommand \Hands[2]{\left({}^\Omega_\tau : {#2} \right)}
\newcommand \HandsP [2]{\left({}^\Omega_\tau : {#1}^{#2} \right)}
\newcommand \HandsPR [4]{\left({}^\Omega_\tau : {#1},{#3}^{#4} \right)}
\newcommand\NewDeal[2]{H^{(#1,#2)}}

\newcommand\TransPlus{{\rm TH}^{d+1}_q}

\newcommand\TransPointPlus[1]{{\rm TH}^{d+1}_q[#1]}
\newcommand\Proj{\pi}
\newcommand\Iota[1]{\iota_{#1}}
\newcommand\ProjDown[1]{\pi^\sigma_{#1}}
\newcommand\ProjUp[1]{\iota^\sigma_{#1}}

\newcommand\deck{\Omega} 
\newcommand\deal[1]{{\Omega\choose #1}}
\newcommand\cardsum[1]{|#1|}
\newcommand\MG{{\sf SP}}
\newcommand\agents{{\mathfrak I}} 
\newcommand\actions{\Lambda}

\newcommand{\vcut}[1]{}

\usepackage[colorinlistoftodos,textwidth=100pt]{todonotes}

\newcommand{\ali}{\ensuremath{\mathcal{A}}\xspace} 
\newcommand{\bob}{\ensuremath{\mathcal{B}}\xspace} 
\newcommand{\cat}{\ensuremath{\mathcal{C}}\xspace}

\newcommand{\dist}{\tau}
\newcommand{\act}{\ensuremath{\alpha}\xspace} 
\newcommand{\run}{\ensuremath{\rho}\xspace} 
\newcommand{\prot}{{\Pi}}

\begin{document}



\title{Perfectly secure data aggregation via shifted projections}


\author{David Fern\'{a}ndez-Duque\\
Department of Mathematics\\
Instituto Tecnol\'ogico Aut\'onomo de M\'exico\\
R\'io Hondo 1, 01080 Mexico City, Mexico\\
{\tt dfduque@us.es}
}

\maketitle

\begin{abstract}
We study a general scenario where confidential information is distributed among a group of agents who wish to share it in such a way that the data becomes common knowledge among them but an eavesdropper intercepting their communications would be unable to obtain any of said data. The information is modelled as a deck of cards dealt among the agents, so that after the information is exchanged, all of the communicating agents must know the entire deal, but the eavesdropper must remain ignorant about who holds each card.

Valentin Goranko and the author previously set up this scenario as the {\em secure aggregation of distributed information} problem and constructed {\em weakly safe} protocols, where given any card $c$, the eavesdropper does not know with certainty which agent holds $c$. Here we present a {\em perfectly safe} protocol, which does not alter the eavesdropper's perceived probability that any given agent holds $c$. In our protocol, one of the communicating agents holds a larger portion of the cards than the rest, but we show how for infinitely many values of $a$, the number of cards may be chosen so that each of the $m$ agents holds more than $a$ cards and less than $2m^2a$.

\end{abstract}

% \linenumbers

\section{Introduction}

Consider a multi-agent network, where each individual holds private information which must be shared among the group, perhaps to reach a consensus or to share a secret. Communication among the agents may be intercepted, leading to the risk of an eavesdropper obtaining confidential data. If encryption is either impossible or undesirable, the agents may use an unconditionally secure protocol, where the exchange would not contain enough information for an eavesdropper to learn a secret \cite{maurer:1999}. This scenario may crop up in many applications, for example when fusing sensorial information or data from computations performed by the individual agents. Alternately, it may have been distributed among the agents in such a way that only by pooling together their knowledge will they have access to sensitive information, as may be the case in {\em secret-sharing protocols} \cite{Blakley1979,shamir1979}.

We will model a situation of this form, where the information is represented by a deck of cards $\deck$ dealt among $m$ agents. The dealing phase is treated as a black box and assumed to be secure. Each of the agents may see her hand, but not the others'. They then want to inform each other of which cards they hold. Meanwhile, the eavesdropper, Eve, may intercept all communications, and the agents do not want her to obtain information about who holds any card. In this setting, we will show that for many possible distributions of cards among the agents, it is indeed possible for them to share the data securely.

\subsection{Comparison to known results}

The model we consider is a multi-agent variation of the well-known {\em Russian cards problem.} The latter may be traced back to \cite{kirkman:1847} but has recently received renewed attention \cite{hvd.studlog:2003}, leading to many new solutions (e.g. \cite{albertetal:2005,geometric,swanson:2012}). In the original Russian cards problem, there are only two communicating agents. In \cite{sadi}, this was generalized by allowing an arbitrary number of agents, but also simplified by assuming that the eavesdropper has no cards in her hand. A key difference between the two-agent and multi-agent setting is that with only two agents, two announcements are usually sufficient for the information exchange, whereas in our setting one might expect to have at least one announcement per agent. However, we remark that longer protocols are already needed to solve some instances with two agents \cite{colouring,threesteps}.

There is more than one way to model the safety constraint. For any card $c$ not held by Eve, she should not know with certainty which agent holds $c$; this is known as {\em weak safety.} But it may be the case that Eve has a very high probability of guessing correctly who holds $c$. To this end, \cite{swanson:2012} introduced the stronger notion of {\em perfect safety,} where Eve's perceived probability that an agent holds $c$ does not change after executing the protocol. Perfectly safe solutions for a wider number of cases were later reported in \cite{swanson2014additional}, and \cite{Landerreche} proposed an approximate notion which led to `almost-perfectly' safe solutions.

In \cite{sadi} we formalized the secure aggregation of distributed information problem and constructed weakly safe solutions for any number of agents. Our goal now is to construct, instead, perfectly safe solutions. These are based on finite linear algebra, and typically one agent holds a large portion of the cards. However, for infinitely many values of $a$, the size of the deck may be chosen so that each of the $m$ agents holds more than $a$ and less than $4m^2 a$ of the cards.

\subsection{Layout of the article}

In Section \ref{SecExample} we present a motivating example illustrating our protocol in an informal setting. Section \ref{sec.SADI} then formalizes the {\em secure aggregation of distributed information} problem and Section \ref{SecPerfect} introduces the notion of perfect safety. Our protocol is based on finite linear algebra, which we review in Section \ref{SecFinGeo} along with some general lemmas we need. In Section \ref{SecProt} we define the protocol and show that it is informative, while in Section \ref{SecSafe} we prove that it is perfectly safe. Finally, in Section \ref{SecParam} we show how one can find relatively balanced card distributions to which the protocol may be applied.

\section{A motivating example}\label{SecExample}

Let us begin by presenting a solution for a distribution of type $(12,2,2)$. This means that Alice draws twelve cards from a deck of sixteen, while Bob and Cath each draw two cards. Let us use $H_\ali,H_\bob$ and $H_\cat$ to denote the set of hands held by Alice, Bob and Cath, respectively. Then, for any card $c$, the probability that a given agent holds $c$ is proportional to the number of cards in their hand; thus, $\prob(c\in H_\ali)=\nicefrac {12}{16}$ and $\prob(c\in H_\bob)= \prob(c\in H_\cat)=\nicefrac {2}{16}$.

Now, let us show how Alice, Bob and Cath can communicate their cards to each other by way of public broadcasts, in such a way that, after the exchange, each of Alice, Bob and Cath knows the entire deal, without changing Eve's perceived probabilities that a given agent holds a given card. Alice will make the first announcement, followed by Bob; it is not necessary for Cath to make an announcement in this setting since she merely holds the complement of Alice and Bob's hands.

\subsection{Alice's announcement}

Here we use the fact that there is a field $\mathbb F_4$, whose elements are $\{0,1,\varphi,\varphi^2\}$ satisfying $1+1=0$ and $\varphi^2=\varphi+1$. With this we construct a 16-point plane, $\mathbb F^2_4$. Alice makes her announcement as follows. First, she randomly assigns each card in the deck to a point on the plane, with the only condition that the cards she {\em does not\footnote{Compare this to the protocol presented in \cite{geometric}, where it is Alice's hand that would form a line rather than its complement.}} hold form a line $\ell$. In the figure, Bob holds spades, Cath holds clubs and Alice holds diamonds, so that $\ell$ is the diagonal $y=x$.

Alice then announces,
\begin{quote}
{\em The complement of my hand forms a line of the form $y=ax+b$ on the plane $\mathbb F^2_4$.}
\end{quote}
Observe that, since there are four choices for each of $a$ and $b$, there are a total of $16$ possible deals according to Alice's announcement.

Because Bob holds two cards, and two points define a line, Bob knows exactly which line his and Cath's cards lie on and thus he knows Alice's hand. Similarly, Cath knows Alice's hand, and in this example in fact Bob and Cath know the entire deal. However, they must inform Alice of their hand. For this, Bob must make an additional announcement.

\begin{figure}[h]
\begin{center}
\begin{tikzpicture}[scale=0.7]
\node (0) at (-1,0) {$0^{\phantom{2}}$};
\node (0) at (-1,1) {$1^{\phantom{2}}$};
\node (0) at (-1,2) {$\varphi^{\phantom{2}}$};
\node (0) at (-1,3) {$\varphi^2$};
\node (0) at (0,-1) {$0^{\phantom{2}}$};
\node (0) at (1,-1) {$1^{\phantom{2}}$};
\node (0) at (2,-1) {$\varphi^{\phantom{2}}$};
\node (0) at (3,-1) {$\varphi^2$};
\node (100) at (0,0) {$\clubsuit$};
\node (101) at (0,1) {$\diamondsuit$};
\node (102) at (0,2) {$\diamondsuit$};
\node (103) at (0,3) {$\diamondsuit$};
\node (110) at (1,0) {$\diamondsuit$};
\node (111) at (1,1) {$\spadesuit$};
\node (112) at (1,2) {$\diamondsuit$};
\node (113) at (1,3) {$\diamondsuit$};
\node (120) at (2,0) {$\diamondsuit$};
\node (121) at (2,1) {$\diamondsuit$};
\node (122) at (2,2) {$\clubsuit$};
\node (123) at (2,3) {$\diamondsuit$};
\node (130) at (3,0) {$\diamondsuit$};
\node (131) at (3,1) {$\diamondsuit$};
\node (132) at (3,2) {$\diamondsuit$};
\node (133) at (3,3) {$\spadesuit$};
\end{tikzpicture}
\end{center}
\caption{Alice holds the complement of the line $y=x$ in the 16-point plane.}
\label{projplane}
\end{figure}
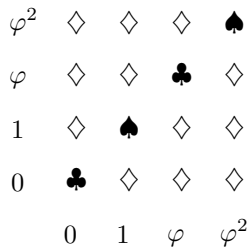

\subsection{A weakly card-safe announcement for Bob}

So it remains for Bob and Cath to inform Alice of their hands. A simple idea that they could use is simply to announce the first coordinates of the points on their hand. In the figure, Bob would say ``The first coordinates of my hand are $\{1,\varphi^2\}$'', while Cath would say ``The first coordinates of my hand are $\{0,\varphi\}$.'' In fact, Cath's announcement follows from Bob's, so she may actually omit it.

As mentioned, Bob and Cath already knew the entire deal, but now Alice can use the fact that each point of $\ell$ is uniquely determined by its first coordinate to infer Bob's and Cath's actual hand. Meanwhile, given any card $c$, Eve cannot tell whether or not Alice holds $c$. Consider, for example, the card $(0,0)$. In the actual deal, Alice does not hold this card, but the only information that Eve has is that Alice holds the complement of a line of the form $y=ax+b$. Thus if $b\not=0$, Alice would in fact hold $(0,0)$. This means that there are three possible values for $b$ which would let Alice hold the origin, and for each choice there are four possible values of $a$, giving us a total of $12$ possible deals (from Eve's perspective) under which Alice holds $(0,0)$.

However, this exchange is not perfectly safe. For indeed, since she knows that Bob's hand projects onto $\{1,\varphi^2\}$, Eve can infer that Bob does not hold the card $(0,0)$. Thus this protocol is only weakly safe. But there is a variation which does yield perfect safety.

\subsection{A perfectly card-safe announcement for Bob}

The only thing that Eve knows about $\ell$ is that it is of the form $y=ax+b$. This $a$ will be used to shift the elements of $\mathbb F_4$ around and add more uncertainty to Bob's announcement. To be precise, if $\ell$ has slope $a$, Bob will now announce, instead of the first coordinates of his hand, these coordinates {\em plus $a$.} Let $\sigma(\ell)$ denote the slope of $\ell$. In this example, since Bob's hands project onto $\{1,\varphi^2\}$ and $\ell$ has slope $1$, Bob will instead announce the value of
\[\{1,\varphi^2\}+\sigma(\ell)=\{1,\varphi^2\}+1=\{0,\varphi\};\]
note that we are using the convention that $X+y=\{x+y:x\in X\}$. His announcement will then take the form
\begin{quote}
``If my hand, $H_\mathcal B$, lies on a line $\ell$, then
\[H_\mathcal B+\sigma(\ell)=\{0,\varphi\}.\text{''}\]
\end{quote}

This new protocol is actually perfectly safe. To see this, first note as before that Alice can hold any card (just replace $\ell$ by a parallel line as needed). But any card may also be held by Bob or by Cath. Let us show this with the card $(0,0)$, held by Cath. Because in the actual deal Cath holds $(0,0)$, Eve evidently finds is possible that Cath holds it.

However, Eve also considers it possible that Bob holds it instead. There are four lines $\ell'$ passing through $(0,0)$ of the form $y=ax+b$, and Eve considers all of these as the possible set of cards that Bob and Cath hold. If $\ell'$ has slope $1$, as is actually the case, then Cath would hold $(0,0)$. But suppose instead that $\ell'$ is the line given by $y=\varphi x$, so that
\[\ell'=\{(0,0),(1,\varphi),(\varphi,\varphi^2),(\varphi^2,1)\}.\]
Then, $\ell'$ has slope $\varphi$, and $\sigma(\ell')=\varphi$. If indeed Bob and Cath's hands lied on $\ell'$, according to Bob's announcement, the first coordinates of his hand would be $0-\sigma(\ell')=\varphi$ and $\varphi-\sigma(\ell')=0$, so that in this case Bob would hold $(0,0)$.

\begin{figure}[h]
\begin{center}
\begin{tikzpicture}[scale=0.7]
\node (0) at (-1,0) {$0^{\phantom{2}}$};
\node (0) at (-1,1) {$1^{\phantom{2}}$};
\node (0) at (-1,2) {$\varphi^{\phantom{2}}$};
\node (0) at (-1,3) {$\varphi^2$};
\node (0) at (0,-1) {$0^{\phantom{2}}$};
\node (0) at (1,-1) {$1^{\phantom{2}}$};
\node (0) at (2,-1) {$\varphi^{\phantom{2}}$};
\node (0) at (3,-1) {$\varphi^2$};
\node (100) at (0,0) {$\spadesuit$};
\node (101) at (0,1) {$\diamondsuit$};
\node (102) at (0,2) {$\diamondsuit$};
\node (103) at (0,3) {$\diamondsuit$};
\node (110) at (1,0) {$\diamondsuit$};
\node (111) at (1,1) {$\diamondsuit$};
\node (112) at (1,2) {$\clubsuit$};
\node (113) at (1,3) {$\diamondsuit$};
\node (120) at (2,0) {$\diamondsuit$};
\node (121) at (2,1) {$\diamondsuit$};
\node (122) at (2,2) {$\diamondsuit$};
\node (123) at (2,3) {$\spadesuit$};
\node (130) at (3,0) {$\diamondsuit$};
\node (131) at (3,1) {$\clubsuit$};
\node (132) at (3,2) {$\diamondsuit$};
\node (133) at (3,3) {$\diamondsuit$};
\end{tikzpicture}
\end{center}
\caption{According to Bob's announcement, if Bob and Cath held the line $y=\varphi x$, this is how the cards would be dealt.}
\end{figure}

More generally, if $\ell'$ is given by $y=ax$, then Bob would hold $(0,0)$ if $a\in \{0,\varphi\}$, while Cath would hold it if $a\in\{1,\varphi^2\}$. This means that, from Eve's perspective, there are two possible deals where Bob holds $(0,0)$ and two possible deals where Cath holds it. But as we saw above, there are $12$ possible deals where Alice holds $(0,0)$; we also saw that there are a total of $16$ possible deals, so Cath's perceived probability that Alice, Bob or Cath holds $(0,0)$ are $\nicefrac{12}{16}, \nicefrac 2{16}$ and $\nicefrac 2{16}$, respectively; the same as before the protocol began!

A similar exercise may be applied to any point $(x,y)$ on the plane to verify that this occurs for any card and thus the protocol is indeed perfectly safe.

\subsection{Generalizing to more agents}

In a more general setting, there may be $m+1$ communicating agents
\[\{\ali,\bob_1,\hdots,\bob_m\}.\]
We will work in $\mathbb F_q^{d+1}$, where $q > m$ is a prime power and $d>0$. Alice will hold all cards except for those that would make up a hyperplane; that is, she would hold $q^{d+1}-q^{d}$ cards, which is more than any other agent. The rest of the agents share the remaining $q^{d}$ cards, with the only restriction that each of them holds more than $q^{d-1}$ of them. Then, Alice will assign a point on $\mathbb F^{d+1}_q$ to each card in such a way complement of her cards form a hyperplane $V$ of the form
\[x_{d+1}=a_1x_1+\hdots +a_dx_d+b\]
and announce her chosen assignment. Each other agent $\bob_k$ holds enough cards to be able to identify $V$, hence Alice's hand. Then, we define
\[\sigma(V)=(a_1,\hdots,a_d)\in\mathbb F^d_q\]
(the `slope' of $V$, similar to a gradient). In our example, the line $\ell$ plays the role of the hyperplane $V$ (since a line on the plane is a hyperplane).

If we denote by $\Proj\colon \mathbb F^{d+1}_q\to \mathbb F^d_q$ the projection onto the first $d$ components, each agent $\bob_k$ will announce\footnote{In general, if $f$ is a function we use $f(x)$ to denote the value of $f$ at a point and $f[X]$ to denote the image of a set $X$ under $f$.} $\Proj[Y_i]+\sigma(V)$, where $Y_i$ is the set of points $\bob_k$ holds (technically, the image of $\bob_k$'s hand under Alice's chosen assignment). As we shall see, even in this more general setting, this protocol will always be informative and perfectly safe. In the remainder of this paper, we will make this precise.

%%%%%%%%%%%%%%%%%%%%%%%%%%%%%%%%
\section{Formalizing the problem} 
 \label{sec.SADI}
%%%%%%%%%%%%%%%%%%%%%%%%%%%%%%%%

Here we will give the basic definitions needed to set up the secure aggregation of distributed information problem, including the notions of informativity and safety that concern us. This section is essentially a review of notions from \cite{sadi}, although we remark that some of the terminology has changed.

%%%%%%%%%%%%%%%%%%%%%%%%%%%%%%%%
\subsection{Basic terminology and notation}
\label{SubsecAnnRunProt}
%%%%%%%%%%%%%%%%%%%%%%%%%%%%%%%%

\begin{definition}
Let $\agents$ be a finite set representing a group of agents. By a {\em distribution type} we mean a vector $\dist =(\dist_\agP)_{\agP\in \agents}$ of positive integers. We write $\cardsum {\dist}$ for $\sum_{\agP\in\agents}\dist_\agP$.

The deck, $\deck$, is a finite set of cards with cardinality $\cardsum {\dist}$. A  {\em deal of type $\dist$ over $\deck$} is a partition $H=(H_\agP)_{\agP\in \agents}$ of $\deck$ such that $|H_\agP|=\dist_\agP$ for each agent $\agP$. We say $H_\agP$ is the {\em hand} of $\agP$. We denote the set of all deals of type  $\dist$  over $\deck$ by $\deal{\dist}$.
\end{definition}

We assume an initial secure dealing phase in which a card deal is selected randomly. Afterwards, the agents have knowledge of their own hand and of the distribution type $\dist$ of the deal, but know nothing more about others' cards. Thus, they are not able to distinguish between different deals where they hold the same hand. We model this by equivalence relations between deals; since from the perspective of agent $\agP$, a deal $H$ is indistinguishable from deal $H'$ whenever $H_\agP=H'_\agP$, we define $H\sim_\agP H'$ if and only if $H_\agP=H'_\agP$. If the agents are numbered $\agP_0,\hdots,\agP_m$, we may write $\sim_k$ instead of $\sim_{\agP_k}$.

We will fix a set $\actions$ representing a language which the agents use to encode information. In a practical setting, elements of $\actions$ would be strings of symbols, but could also be modelled as natural numbers; we will refer to them simply as {\em tokens.} For simplicity we will assume that agents take turns, so that if they are listed by $\agP_0,\hdots,\agP_m$, then $\agP_0$ places a token first, followed by $\agP_2$, etc.

\begin{definition}[Run]\label{def:run}
Let $\actions$ be a set whose elements will be called {\em tokens.} A {\em (finite) run} is a (possibly empty) sequence $\run=\act_0,\hdots,\act_n$ of tokens from $\actions$. The empty run is denoted by $()$. If $\run=\act_0,\hdots,\act_n$ and $\act$ is a token we write $\rho \ast \act$ for $\act_0,\hdots,\act_n,\act$. An {\em infinite run} is an infinite sequence $\act_0,\act_1,\act_2,\hdots$ of tokens. Runs will be assumed finite unless it is explicitly stated otherwise. We denote the length of a run $\rho$ by $|\rho|$. We denote the set of finite runs by ${\rm Run}$.
\end{definition}

We now define the notion of {\em protocol} we will use. Below we use $(x)_{d}$ to mean the remainder of $x$ modulo $d$.

\begin{definition}[protocol]\label{defprot}
Let $\dist$ be a distribution type over $\agents=(\agP_0,\hdots,\agP_m)$. A {\em protocol} (for $\dist$) is a function $\prot$ assigning to every deal $H\in \deal{\dist}$ and every run $\rho\in{\rm Run}$ a set of tokens $\prot(H,\rho)\subset \actions$ such that if $k=(|\rho|)_{m+1}$ (so that it is the turn of the agent $\agP_k$) and $H\sim_k H'$, then $\prot(H,\rho)=\prot(H', \rho)$.
\end{definition}

Thus, a protocol is a tree-like set of runs representing a non-deterministic protocol for the communicating agents. Once a deal has been fixed, a protocol assigns to each run a set of tokens out of which the agent whose turn it is must choose one at random. These tokens are determined exclusively by the information the agent has access to, which is assumed to be {\em only:} ({\it i}) her hand, ({\it ii}) the distribution type  $\dist$ and the deck $\deck$, ({\it iii}) the announcements that have been made previously and ({\it iv}) the protocol being executed.

Note that protocols are generally non-deterministic and hence may have many {\em executions:}

\begin{definition}
An {\em execution of a protocol $\prot$} is a pair $(H,\run)$ consisting of a deal $H\in {\deal{\dist}}$ and a run $\run=\act_0,\hdots,\act_n$, such that $\act_{k} \in \prot(H,\rho_{<k})$ for every $k\leq n$, where $\rho_{<k} = \act_0,\hdots,\act_{k-1}$ ($\rho_{<0}$ is empty). We say that $\rho$ is a {\em run of $\Pi$} if there exists a deal $H$ such that $(H,\rho)$ is an execution of $\Pi$.

An execution of a protocol $(H,\run)$ is {\em terminal} if $\prot(H,\run)=\varnothing$. A protocol is {\em terminating} if it has no infinite executions.
\end{definition}

%%%%%%%%%%%%%%%%%%%%%%%%%%%%%%%%
\subsection{Informative and weakly safe protocols}
\label{SubsecTypesAnn}
%%%%%%%%%%%%%%%%%%%%%%%%%%%%%%%%

Now we will define some desirable properties that protocols may have. The first property is \emph{informativity}: that agents in the team learn the entire deal at the end of its execution:

\begin{definition}[Informativity]
An execution $(H,\run)$ of a protocol $\prot$ is  {\em informative for an agent $\agP$} if there is no execution $(H' ,\run)$ of $\prot$ with $H'\not=H$ but $H_\agP=H'_\agP$ (i.e., at the end of the run the agent knows the precise card distribution.) 

A terminating protocol $\prot$ is {\em informative} if every terminating execution of $\prot$ is informative for \emph{every} agent in $\agents$.
\end{definition}

In \cite{sadi} we considered a weak notion of safety and showed that informative and weakly safe protocols exist for a large class of distribution types.

\begin{definition}[Weak safety of protocols]\label{def:cardsafe}
An execution $(H,\run)$ of a protocol $\prot$ is {\em weakly safe for the card $c$} if there are agents $\agP\not =\agQ$ and a deal $H'\in \deal\dist$ such that $(H',\rho)$ is also an execution of $\prot$ but $c\in H_\agP$ and $c\in H'_\agQ$.

A protocol $\prot$ is weakly safe if every execution of $\prot$ is weakly safe for every card.
\end{definition}

However, a weakly safe protocol may give Eve a large amount of probabilistic information. To this effect, in the next section we will turn to formalizing a stronger notion of safety.

\section{Perfect safety}\label{SecPerfect}

A weakly safe protocol does not allow Eve to know with certainty who holds a given card $c$, but she may gain other information about it; for example, she may learn that a certain agent $\agP$ {\em does not} hold $c$, or perhaps that an agent $\agQ$ is very likely to hold $c$. Thus it will be desirable to control the probabilistic information that Eve obtains.

In order to do so, we will make two assumptions:
\begin{enumerate}
\item The dealer chooses uniformly from the set of all deals.

\item In any protocol $\Pi$, each agent always chooses uniformly from $\Pi(H,\rho)$, provided it is non-empty.

\end{enumerate}
To compute the relevant probabilities, it will be useful to introduce the notation $\left ({}^\Omega_\tau : C_1,\hdots,C_n \right )$ to denote the set of deals satisfying the constraints $C_1,\hdots,C_n$.

\begin{definition}\label{DefHands}
Fix a deck $\deck$, a distribution type $\dist$ and a protocol $\prot$. Then define:
\begin{itemize}
\item For a run $\rho$, $\Hands \prot\rho$ to be the set of all $H\in{\Omega\choose{\dist}}$ such that $(H,\rho)$ is an execution of $\Pi$. Elements of $\Hands \prot\rho$ are {\em possible deals} (according to Eve).
\item For a card $c$ and an agent $\agP$, $\HandsP c \agP$ to be the set of deals $H$ such that $c\in H_\agP$.
\item For a run $\rho$, a card $c$ and an agent $\agP$,
\[\HandsPR \run\prot c\agP=\Hands \prot\rho\cap\HandsP c\agP,\]
the set of possible deals where $\agP$ holds $c$.
\end{itemize} 
\end{definition}

With this we can define the following generalization of a notion introduced in \cite{swanson:2012}.

\begin{definition}[Equitative protocol]\label{DefEquit}
A protocol $\Pi$ is {\em equitative} if for every run $\rho$ of $\prot$ there is a constant $k=k(\run)$ such that for every deal $H\in \Hands\prot\rho$ we have that $|\Pi(H,\rho)|=k$.
\end{definition}

In other words, $|\Pi(H,\rho)|$ depends on $\run$ but not on $H$. Equitative protocols will allow us to simplify many computations. Throughout the text, we use $\prob$ to denote probability.

\begin{lemm}\label{LemmCompBasicProb}
Let $\prot$ be an equitative protocol some distribution type $\dist$ and $\rho$ be a run of $\prot$. Then, there is a constant $\gamma=\gamma(\run)\in(0,1]$ so that, for any deal $H$,
\[
\prob \left (\rho \mid H \right )
=
\begin{cases}
\gamma&\text{if $H\in\Hands\prot\run$}\\
0&\text{otherwise.}
\end{cases}
\]
\end{lemm}

\proof
It is obvious that $\prob\left (\rho \mid H\right )=0$ if $H\not\in\Hands\prot\rho$. Otherwise, we proceed by induction of the length of $\run$.

For the base case, $\run=()$ and $\prob\left (\rho \mid H\right )=1$ (since every execution begins with the empty run). Otherwise, we consider a run of the form $\run\ast\act$. Let $k=k(\rho)$ be such that $|\prot(H,\run)|=k$ for any deal $H\in \prot(H,\run)$. By induction hypothesis, $\prob\left(\rho\mid H\right )=\gamma'=\gamma'(\rho)$ for any $H\in \Hands \prot{\rho}$. Then, for any deal $H\in \Hands\prot\run$,
\[\prob \left (\rho\ast\act\mid H\right ) =\prob\left (\act\mid \rho,H\right) \, \prob \left(\rho\mid H\right)= (\nicefrac 1k) \gamma' \]
and we can set $\gamma(\run\ast \act)=\nicefrac{\gamma'}k$.
\endproof

\begin{prop}\label{PropComputeProb}
Let $\Pi$ be an equitative protocol over some distribution type $\dist$. Then, for any run of the protocol $\rho$, $c\in\deck$ and any $\agP\in\agents$,
\[\prob\left (c\in H_\agP \mid \rho \right)=\dfrac{\left |\HandsPR \run\prot c\agP \right|}{\left | \Hands\prot\rho \right |}.\]
\end{prop}

\proof
By Bayes' law,
\begin{align}
\nonumber \prob \left (c\in H_\agP \mid \rho \right )&=\dfrac{\prob \left (c\in H_\agP , \rho\right)}{\prob (\rho)}\\
 &=\dfrac{\displaystyle\sum_{H\in \HandsP c\agP} \prob \left (\rho \mid H \right ) \prob(H)}{\displaystyle\sum_{H\in\deal\tau} \prob \left ( \rho \mid H  \right ) \prob (H) }.\label{EqBayes}
\end{align}
But by Lemma \ref{LemmCompBasicProb}, $\prob \left ( \rho \mid H \right )$ is some constant $\gamma=\gamma(\run)$ if $H\in \Hands\prot\run$ and zero otherwise. Moreover, since deals are also chosen uniformly, $\prob (H)=\delta$ for some constant $\delta$.

Thus, \eqref{EqBayes} becomes
\begin{align*}
\dfrac{\displaystyle\sum_{H\in \HandsPR \run\prot c\agP  } \prob \left (\rho\mid H \right) \prob(H) }{\displaystyle\sum_{H\in\Hands\prot \rho}  \prob\left ( \rho \mid H \right ) \prob (H) }&=\dfrac{\displaystyle \sum_{H\in \HandsPR \run\prot c\agP  } \gamma\delta }{\displaystyle\sum_{H\in\Hands\prot \rho} \gamma\delta }\\
&=\dfrac{\left | \HandsPR \run\prot c\agP  \right | }{|\Hands \prot\rho |},
\end{align*}
as claimed.
\endproof

\section{Geometric preliminaries}\label{SecFinGeo}

Our perfectly safe solution is based on finite linear algebra. We assume some basic familiarity with finite fields and finite geometry; these are covered in texts such as \cite{lidl1997} and \cite{dembowski1997}, respectively.

Throughout the paper, $q$ will denote a prime or a power of a prime, and $\mathbb F_q$ the field with $q$ elements. If $d$ is any natural number, $\mathbb F^d_q$ denotes the vector space of dimension $d$ over $\mathbb F_q$. Given $U\subset\mathbb F^d_q$ and $v\in\mathbb F^d_q$, we write $U+ v$ for the set $\{u+v:u\in U\}$. A {\em hyperspace} is a subspace of dimension $d-1$, and by a {\em hyperplane} we mean any set of the form $V+x$, where $V$ is a hyperspace. Two hyperplanes $X,Y$ are {\em parallel} if $X\not=Y$ but there is a vector $x$ such that $X=Y+x$.
 
Recall that $|\mathbb F^d_q|=q^d$, where in general $|X|$ denotes the cardinality of $X$. Moreover, if $U\not =V$ are hyperplanes, then $U$ has exactly $q^{d-1}$ elements, while $|U\cap V|\leq q^{d-2}$ and equality holds unless $U,V$ are parallel, in which case their intersection is empty.

In our example in Section \ref{SecExample}, it was important that the line $\ell$ have an equation of the form $y=ax+b$. This readily generalizes to the higher-dimensional setting, as defined below.

\begin{definition}
Given a prime power $q$ and $d>0$, we say that $V\subset\mathbb F^{d+1}_q$ is a {\em transversal hyperplane} if there are $a_1,\hdots,a_d\in \mathbb F_q$ such that $V$ is the graph of
\[x_{d+1}=a_1x_1+\hdots+a_{d}x_{d}+b.\]
If $b=0$ then we say $V$ is a {\em transversal hyperspace.}

We denote the set of transversal hyperplanes in $\mathbb F^{d+1}_q$ by $\TransPlus$. Given $x\in\mathbb F^{d+1}_q$, we denote by $\TransPointPlus x$ the set of transversal hyperplanes in $\mathbb F^{d+1}_q$ touching $x$.
\end{definition}

It will be useful (and straightforward) to count the number of transversal hyperplanes in $\mathbb F^{d+1}_q$.

\begin{lemm}\label{LemmCountTrans}
Fix a prime power $q$ and a natural number $d$. Then,
\begin{enumerate}
\item $\left |\TransPlus \right |=q^{d+1}$;
\item for any $x\in \mathbb F^{d+1}_q$, $\left |\TransPointPlus x \right |=q^{d}$.
\end{enumerate}
\end{lemm}

\proof
For the first claim note that a transversal hyperplane is determined by an equation
\[x_{d+1}=a_1x_1\hdots a_dx_d+b\]
and there are $q$ choices for each $a_k$ as well as for $b$, giving a total of $q^{d+1}$ options.

For the second, we may assume that $x=\vec 0$ without loss of generality. This forces us to set $b=0$. Then we must merely choose each $a_k$, and since there are $d$ of them we have $q^d$ options.
\endproof

Next, we define the `slope' of a transversal hyperplane, which is itself a vector.

\begin{definition}
Given a prime power $q$ and a positive integer $d$, we define $\sigma\colon \TransPlus \to \mathbb F^{d}_q$ given by $\sigma(V)=(a_1,\hdots,a_d)$ whenever $a_1,\hdots,a_d\in \mathbb F_q$ are such that $V$ is the graph of
\[x_{d+1}=a_1x_1+\hdots+a_{d}x_{d}+b.\]
\end{definition}

The following is then immediate:

\begin{lemm}\label{LemmSigma}
Given a prime power $q$, $d>0$ and $x\in\mathbb F^{d+1}_q$, the restriction $\sigma \colon\TransPointPlus x \to \mathbb F^d_q$ is a bijection.
\end{lemm}

It will also be useful to project the points on a transversal hyperplane $V$ onto their first $d$ coordinates. In particular, the projection of one such point will be sufficient to determine its missing component, provided we know what $V$ is.

\begin{definition}
Fix a prime power $q$ and $d>0$. We define $\Proj\colon \mathbb F^{d+1}_q\to \mathbb F^{d}_q$ by $\Proj(x_1,\hdots,x_{d+1})=(x_1,\hdots,x_{d}).$
\end{definition}

\begin{lemm}\label{LemmIota}
Given a prime power $q$, $d>0$ and $V\in \TransPlus$, $\Proj\colon V\to \mathbb F^{d}_q$ is a bijection. We denote its inverse by $\Iota V$.
\end{lemm}

\proof
If $V$ is given by the equation
\[x_{d+1}=a_1x_1 + \hdots + a_dx_d + b,\]
then it is easy to see that $\pi\colon V\to \mathbb F^d_q$ has an inverse given by
\[\Iota V (x_1,\hdots, x_d) = (x_1,\hdots, x_d, a_1x_1 + \hdots + a_dx_d + b ).\qedhere\]
\endproof

In Section \ref{SecExample}, we saw that Bob constructed his announcement by projecting and shifting. We also saw that Alice could reconstruct Bob's hand from this announcement. Let us now present these operations in more generality.

\begin{definition}
Let $V$ be a transversal hyperplane of $\mathbb F^{d+1}_q$. We define $\ProjDown V\colon V \to \mathbb F^d_q$ by
\[\ProjDown V (w)=\Proj(w)+\sigma(V)\]
and $\ProjUp V \colon \mathbb F^d_q \to V$ by
\[\ProjUp V (y)=\Iota V \big (y-\sigma(V) \big ).\]
\end{definition}

\begin{lemm}\label{LemmInverse}
If $q$ is a prime power, $d>1$ and $V\in\TransPlus$, then $\ProjDown V\circ \ProjUp V$ is the identity on $\mathbb F^d_q$ and $\ProjUp V\circ \ProjDown V$ is the identity on $V$.
\end{lemm}

\proof
Using Lemma \ref{LemmIota}, we see that, for $x\in \mathbb F^d_q$,
\begin{align*}
\ProjDown V\circ \ProjUp V(x)&=\ProjDown V(\Iota V(x-\sigma(V)))=\Proj(\Iota V(x-\sigma(V)))+\sigma(V)\\
&=x-\sigma(V) +\sigma(V)=x.
\end{align*}
That $\ProjUp V\circ \ProjDown V$ is the identity on $V$ is proven similarly.
\endproof

\section{The {\protname} protocol}\label{SecProt}

With these ingredients we are ready to define our protocol. It depends on several parameters which must be `suitably' chosen, in the following sense.

\begin{definition}[suitable parameters]
We say $(m,q,d,\dist)$ are {\em suitable parameters} if $m> 1$, $q > m$ is a prime power, $d>0$, and $\dist$ is a distribution type over $\agents=\{\ali,\bob_1,\hdots,\bob_m\}$ such that $|\dist|=q^{d+1}$, $\dist_\ali=q^{d+1}-q^{d}$ and, for each $k\in[1,m]$, $\dist_{\bob_k}>p^{d-1}$.
\end{definition}

Once we have selected suitable parameters, our protocol may be fully determined by describing its maximal executions, since all other executions will merely be initial segments of these. We will use this idea in order to simplify the following definition. 

\begin{definition}[{\protname} protocol]
Let $(m,q,d,\dist)$ be suitable parameters and $\deck$ be any set with $q^d$ elements. Then, given a deal $H\in{\Omega\choose {\dist}}$, the maximal executions of the {\protname} protocol are of the form
\[ (H,f,X_1,\hdots,X_m),\]
where
\begin{itemize}
\item $f\colon \deck\to \mathbb F_q^d$ is such that $V=\mathbb F^{d+1}_q\setminus f[H_\ali]$ is a transversal hyperplane and

\item for each $k\in [1,m]$, $X_k= \ProjDown V[f[H_{\bob_k}]]$.
\end{itemize}
The {\protname} protocol will be denoted $\MG$.
\end{definition}

Our goal is to prove the following:

\begin{theorem}\label{TheoMain}
The {\protname} protocol is informative and perfectly safe for any choice of suitable parameters.
\end{theorem}

Before proceeding, we must check that we have actually given a protocol according to our definitions.

\begin{lemm}\label{LemmIsProt}
Given any choice of suitable parameters, the {\protname} protocol is an equitative protocol.
\end{lemm}

\proof
We begin by checking that our protocol satisfies Definition \ref{defprot}. Suppose that $(m,q,d,\dist)$ are suitable parameters. Let $(H,\rho)$ be an execution of the {\protname} protocol and $\act\in\MG(H,\run)$. Let $\agP$ be the last agent to make an announcement and $H'$ be a deal such that $H_\agP=H'_\agP$ and $(H,\run)$ is an execution of our protocol. We must check that $\act\in\MG(H',\run)$ as well.

First assume that $\rho$ is empty, so that $\act$ is Alice's announcement of $f\colon \deck\to\mathbb F^{d+1}_q$. Then, $\mathbb F^{d+1}_q \setminus f[H_\ali] = \mathbb F^{d+1}_q \setminus f[H'_\ali]$, so that $V=\mathbb F^{d+1}_q \setminus f[H'_\ali]$ is a transversal hyperplane, and since $f$ was already a bijection, $f\in\MG \big ( H',()\big )$.

Otherwise, $\agP=\bob_k$ for some $k$, and the last announcement is of the form $X_k=\ProjDown V[f[H_{\bob_k}]]$. Let $V'=\mathbb F^{d+1}_q\setminus f[H'_\ali]$. Then, $V'$ is a hyperplane containing $H'_{\bob_k}=H_{\bob_k}$. Thus, $H_{\bob_k}\subseteq V\cap V'$ and hence $|H_{\bob_k}|\leq |V\cap V'|$. But if $V\not=V'$ then $|V\cap V'|\leq q^{d-1}<|H_{\bob_k}|$, which is impossible. We conclude that $V=V'$ and thus $X_k=\ProjDown {V'} [f[H'_{\bob_k}]]$. It follows that $X_k\in\MG \big ( H',\run )$.

It remains to check that the protocol is equitative in the sense of Definition \ref{DefEquit}, that is, that $|\MG(H,\rho)|$ depends on $\rho$ and not on $H$. This is not hard to see: when $\rho=()$, the number of bijections $f\colon \deck\to \mathbb F^{d+1}_q$ such that $f[H_\ali]$ is the complement of a hyperplane clearly does not depend on $H$, since different deals are obtained merely by permuting the cards. If on the other hand $\rho=f,X_1,\hdots,X_{k-1}$, then the value of $X_{k}$ is uniquely determined by the expression $X_{k}= \ProjDown V[f[H_{\bob_k}]]$; hence $|\MG(H,\rho)|=1$ for any deal $H$. We conclude that the {\protname} protocol is an equitative protocol, as claimed.
\endproof

Now that we know we have a protocol, let us check that it is indeed informative and perfectly safe. We will proceed by breaking the proof into several steps. First, let us check that the protocol is informative.

\begin{lemm}\label{LemmInf}
The {\protname} protocol is informative for any choice of suitable parameters.
\end{lemm}

\proof
Let $(m,q,d,\dist)$ be suitable parameters. Let $(H,\rho)$ be a terminal execution of the protocol, and let $\agP\not= \agQ$ be agents. We must check that, if $(H',\rho)$ is another terminal execution of the protocol with $H'_\agP=H_\agP$, then also $H'_\agQ=H_\agQ$.

First assume that $\agQ=\ali$, so that $\agP=\bob_j$ for some $j$. In this case, $V=\mathbb F^{d+1}_q\setminus f[H_\ali]$ is the unique hyperplane such that $f[H_{\bob_j}]\subset V$, and similarly $V'=\mathbb F^{d+1}_q\setminus f[H'_\ali]$ is the unique hyperplane such that $f[H_{\bob_j}]=f[H'_{\bob_j}]\subset V'$. It follows that $V= V'$ as well and thus $H_\ali=H'_\ali$.

Now assume that $\agQ=\bob_k\not =\ali$. Note that by the previous case, $H_\ali = H'_\ali$ and thus if we set $V=\mathbb F^{d+1}_q\setminus f[H_\ali]$, then we also have $V=\mathbb F^{d+1}_q \setminus f[H'_\ali]$. It follows by the definition of the protocol that $\bob_k$ has made an announcement of the form $X_k= \ProjDown V [f[H_{\bob_k}]]$, so that by Lemma \ref{LemmInverse}, $\ProjUp V [X_k]=f[H_{\bob_k}]$. Similarly, since $(H',\run)$ is also an execution of our protocol, $\ProjUp V [X_k]=f[H'_{\bob_k}]$. Thus $f[H_{\bob_k}]=f[H_{\bob_k}]$; since $f$ is a bijection, $H_{\bob_k}=H'_{\bob_k}$, as claimed.
\endproof

It remains to check that the {\protname} protocol is perfectly safe. This will require a bit more work.

\section{Perfect safety of the {\protname} protocol}\label{SecSafe}

To prove that the {\protname} protocol is perfectly safe, we will construct new deals that the eavesdropper may consider possible after its execution. The following definition shows how we will do this.

\begin{definition}
Let $(m,q,d,\dist)$ be suitable parameters. Suppose that
\[\rho= f,X_1,\hdots,X_m \]
is such that $f\colon \deck\to \mathbb F^{d+1}_q$ and each $X_i\subset \mathbb F^d_q$, and let $V\in \TransPlus$.

For each agent $\agP$ define a hand $\NewDeal V\rho _\agP$ by
\begin{itemize}
\item $\NewDeal V\rho_\ali=f^{-1}[\mathbb F^{d+1}_q\setminus V]$
\item for $k\in [1,m]$, $\NewDeal V\rho_{\bob_k}=f^{-1} [ \ProjUp V [X
_k] ]$.
\end{itemize}
\end{definition}

\begin{lemm}\label{LemmIsDeal}
Let $(m,q,d,\dist)$ be suitable parameters. If
\[\rho=f,X_1,\hdots,X_m\]
is such that $f\colon \deck\to \mathbb F^{d+1}_q$ is a bijection, $X_1,\hdots,X_m$ form a partition of $\mathbb F^{d}_q$ and $|X_j|=\dist_j$ for all $j$, then for any transversal hyperplane $V$, $\NewDeal V\rho$ is a deal of distribution type $\dist$ and $(\NewDeal V\rho ,\rho)$ is an execution of the {\protname} protocol.
\end{lemm}

\proof
First let us check that $\NewDeal V\rho$ is a deal of distribution type $\dist$. For it to be a deal merely means that it is a partition of $\deck$. Since $|\dist|=q^{d+1}=|\deck|$, this boils down to checking that all hands are disjoint and that each agent $\agP$ holds $\dist_\agP$ cards. So suppose $\agP\not = \agQ$ are two agents. If one of them (say, $\agP$) is Alice and $\agQ=\bob_k$, then we observe that Alice holds the complement of $f^{-1}[V]$ whereas $\ProjUp V [X_k] \subset V$, so that $\NewDeal V\rho _{\bob_k}=f^{-1}[\ProjUp V [X_k]]\subset f^{-1}[V]$ and hence the two agents' hands are disjoint. If on the other hand $\agP=\bob_j$ and $\agQ=\bob_k$, then since $f^{-1}$ and $\ProjUp V$ are injective and $X_j$ and $X_k$ are disjoint, then $\NewDeal V\rho _{\bob_j}=f^{-1} [ \ProjUp V [X_j] ]$ is disjoint from $\NewDeal V\rho_{\bob_k}=f^{-1} [ \ProjUp V [X_k] ]$. We conclude that all hands of $\NewDeal V\rho$ are disjoint.

The injectivity of $f^{-1}$ and $\ProjUp V$ also gives us
\[\left|\NewDeal V\rho _{\ali}\right|=|\mathbb F^{d+1}_q\setminus V|=q^{d+1}-q^d=\dist_\ali,\]
as well as
\[\left | \NewDeal V\rho _{\bob_k}\right |=\left|f^{-1} [ \ProjUp V [X_k] ]\right|=|X_k|=\dist_{\bob_k}\]
for all $k\in[1,m]$, so indeed $\NewDeal V\rho$ is a deal of distribution type $\dist$.

Finally, let us check that $(\NewDeal V\rho,\rho)$ is an execution of the {\protname} protocol. We have assumed that $f$ is bijective and that $f[\NewDeal V\rho_\ali]=\mathbb F^{d+1}_q\setminus V$ is obvious by the definition of $\NewDeal V\rho_\ali$. Meanwhile, using Lemma \ref{LemmInverse}, we have for each agent $\bob_k$ that
\[\ProjDown V\circ f [ \NewDeal V\rho_{\bob_k}]=\ProjDown V\circ f [ f^{-1} \circ \ProjUp V[ X_k]]=\ProjDown V \circ  \ProjUp V[ X_k]=X_k,\]
so that indeed $(\NewDeal V\rho,\rho)$ is an execution of our protocol.
\endproof

Moreover, $\NewDeal V\rho$ is unique, in the following sense.

\begin{lemm}\label{LemmUniqueDeal}
Let $(m,q,d,\dist)$ be suitable parameters. If $H$ is a deal, $\rho=f,X_1,\hdots,X_m$ is a run such that $(H,\rho)$ is an execution of the {\protname} protocol and $V=\mathbb F^{d+1}_q\setminus f[H_\ali]$, then $H=\NewDeal V\rho$.
\end{lemm}

\proof
Once we have fixed $V$, then for any agent $\bob_k$ we must have $H_{\bob_k}= \Iota V [X_k]= \NewDeal V\rho_{\bob_k}$, and since Alice also holds the same hand in $H$ and $\NewDeal V\rho$, the two deals must be equal.
\endproof

The deals $\NewDeal V\rho$ will be essential in showing that the protocol is perfectly safe. In fact, we will show that given any agent $\agP$ and any run of the protocol $\rho$, the set of possible deals where $\agP$ holds $c$ is precisely $\dist_\agP$, as was the case in the example on Section \ref{SecExample}. Below we use the notation $\left ({}^\Omega_\tau : C_1\hdots,C_n \right ) $, introduced in Definition \ref{DefHands}.

\begin{lemm}\label{LemmPreSafe}
Let $(m,q,d,\dist)$ be suitable parameters, $\rho$ be a run of the {\protname} protocol and $\agP$ be any agent. Then,
\[\left| \HandsPR \run\MG c\agP  \right | = \dist_\agP.\]
\end{lemm}

\proof
Suppose that $\rho=f,X_1,\hdots,X_m$ and let $c\in\deck$ be any card. We will consider the cases where $\agP=\ali$ and where $\agP=\bob_k$ separately.

\paragraph{Counting deals where Alice holds $c$} Let $c$ be any card; we wish to count the number of deals $H$ such that $(H,\rho)$ is an execution of the {\protname} protocol and $c\in H_\ali$. Now, the complement of $f[H_\ali]$ is a transversal hyperplane $V$, which should not contain $f(c)$. By Lemma \ref{LemmCountTrans}, there are $q^{d+1}$ transversal hyperplanes and $q^d$ touching $f(c)$, which leaves $q^{d+1}-q^{d}$ avoiding $f(c)$. By Lemma \ref{LemmIsDeal}, for each such $V$, $\NewDeal V\rho$ is a deal such that $(\NewDeal V\rho,\rho)$ is an execution of the {\protname} protocol, and where Alice holds $c$. Moreover, by Lemma \ref{LemmUniqueDeal}, this is the unique deal with such properties. It follows that the possible deals where Alice holds $s$ are in bijection with the set of transversal hyperplanes avoiding $f(c)$, and thus
\[\left| \HandsPR \run\MG c\ali  \right | = q^{d+1}-q^d = \dist_\ali.\]

\paragraph{Counting deals where another agent holds $c$}
Now consider the case where $\agP=\bob_k$ for some $k$. In this case, we claim that the possible deals where $\bob_k$ holds $c$ are in bijection with $X_k$. For this, we will define a function
\[h\colon X_k\to  \HandsPR \run\MG c{\bob_k} \]
and show that it is bijective.

Fix $v\in X_k$. Let $w=\Proj (f(c))$, and pick the unique $U\in\TransPointPlus{f(c)}$ such that $\sigma(U)=v-w$ (which exists by Lemma \ref{LemmSigma}). Denote this $U$ by $U^v$.

Now, let $h(v)=\NewDeal {U^v}\rho$. We claim that $h$ gives the desired bijection. First let us check that $h(v)\in \HandsPR \run\MG c{\bob_k} $ whenever $v\in X_k$. By Lemma \ref{LemmIsDeal}, $h(v)=\NewDeal {U^v}\rho$ is a new deal and $(h(v),\rho)$ is an execution of the {\protname} protocol, so $h(v)\in \Hands \MG\rho$. Moreover, note that
\[\ProjUp {U^v} (v)=\Iota {U^v}(v-\sigma(U^v))=\Iota {U^v}(w)=f(c).\]
But $v\in X_k$ so $f(c)\in \ProjUp {U^v} [X_k]$, that is,
\[c\in f^{-1}[\ProjUp W[X_k]]=\NewDeal {U^v}\rho _{\bob_k}=h(v)_{\bob_k}.\]
In other words, $\bob_k$ holds $c$ in the deal $h(v)$; by definition, this means that $h(v)\in \HandsP c {\bob_k}$. We conclude that $h(v)\in \HandsPR \run\MG c{\bob_k}$, as claimed.

Next let us check that $h$ is injective. If $v\not=v'\in X_i$ then $v-w\not=v'-w$, so that $U^{v}\not= U^{v'}$ and thus $h(v)\not=h(v')$, since Alice would hold a different hand in each deal. Since $v,v'$ were arbitrary, we conclude that $h$ is indeed injective.

Finally, let us see that $h$ is onto. Let $H$ be any deal where $\bob_k$ holds $c$ and such that $(H,\rho)$ is an execution of $\MG$. Let $V$ be the complement of $f[H_\ali]$. Then, $X_k=\ProjDown V [f[H_{\bob_k}]]$, so that $\pi (f(c))+\sigma(V) \in X_k$. As before, let $w=\pi (f(c))$ and $v = w+\sigma(V)$. Then, $v\in X_k$ and $v-w=\sigma(V)$. But since $V$ touches $f(c)$ and $\sigma$ is a bijection when restricted to $\TransPointPlus {f(c)}$ (once again by Lemma \ref{LemmSigma}), it follows that $V=U^v$ and, by Lemma \ref{LemmUniqueDeal}, $H=\NewDeal V\rho= h(v)$. Since $H$ was arbitrary, we conclude that $h$ is onto.

Therefore $h$ is a bijection and
\[\left| \HandsPR \run\MG c{\bob_k}  \right | = |X_k| = \dist_{\bob_k},\]
as desired.

Since we have now considered all possible cases for $\agP\in\agents$, the lemma follows.
\endproof

We now have all the ingredients we need to prove our main theorem.

\begin{lemm}\label{LemmSafe}
The {\protname} protocol is perfectly safe for any choice of suitable parameters.
\end{lemm}

\proof
Let $(m,q,d,\dist)$ be suitable parameters, $c$ a card and $\agP$ an agent. By Lemma \ref{LemmPreSafe}, $\left| \HandsPR \run \MG  c\agP  \right | = \dist_\agP.$ Moreover, $|\Hands\MG\rho|$ is equal to the number of transversal hyperplanes in $\mathbb F^{d+1}_q$, which by Lemma \ref{LemmCountTrans} is $q^{d+1}=|\dist|$. Thus by Proposition \ref{PropComputeProb},
\[\prob (c\in H_\agP | \rho )=\dfrac{\left| \HandsPR \run \MG c\agP  \right |}{\left| \Hands \MG\rho \right |}=\frac{\dist_\agP}{|\dist|},\]
and since $\agP$ was arbitrary, this means that the protocol is perfectly safe.
\endproof

With this, we may prove our main result.

\proof[Proof of Theorem \ref{TheoMain}]
By Lemma \ref{LemmIsProt}, the {\protname} protocol is a protocol according to Definition \ref{defprot}; moreover, by Lemma \ref{LemmInf}, it is informative, whereas by Lemma \ref{LemmSafe}, it is perfectly safe, as needed.
\endproof

\section{Finding balanced distribution types}\label{SecParam}

The {\protname} protocol has the disadvantage that one agent must hold a disproportionate portion of the cards. However, this can be controlled to a certain extent. In this section we will show how, given the number $m$ of agents, one may find suitable distribution types over $m$ agents that are not too unbalanced.

For this we will use the following lemma.

\begin{lemm}\label{LemmTwoM}
Given a natural number $m>0$ there is a prime power $q$ such that $m < q \leq 2m$.
\end{lemm}

\proof
Just take $q$ to be the unique power of $2$ satisfying the required bounds.
\endproof

There are many possible improvements to this result (for example we may take $q$ to be prime using Bertrand's postulate), but this simple version will suffice for our purposes. With this, we may prove the following.

\begin{corollary}\label{CorMain}
Given a set $\agents=\{\ali,\bob_1,\hdots,\bob_m\}$ of $m+1$ agents, there are infinitely many values of $a$ such that the {\protname} protocol is informative and perfectly safe for some distribution type $\dist$ over $\agents$ such that, for each agent $\agP\in\agents$, $ \dist_\agP\in (a, 4m^2 a)$.
\end{corollary}

\proof
Fix $m$ and use Lemma \ref{LemmTwoM} to find a prime power $q\in (m,2m]$. Fix an arbitrary $d>1$ and define $\dist$ by setting $\dist_\ali=q^{d+1}-q^d$ and, for $k\in[1,m-1]$, $\dist_{\bob_k}=q^{d-1}+1$. Finally, let $\dist_{\bob_m}=q^{d}-(m-1)(q^{d-1}+1)$. Set $a=q^{d-1}$.

Clearly $|\dist|=q^{d+1}$, while
\[\dist_\ali=q^{d+1}-q^d<q^{d+1}\leq 4 m^2 q^{d-1}=4m^2 a.\]
For $k\in[1,m-1]$, it is obvious that $\dist_k>q^{d-1}$, while $\dist_m>q^{d-1}$ because
\[q^{d}-(m-1)(q^{d-1}+1)\geq q^{d}-(q-2)(q^{d-1}+1)=2q^{d-1}-q+2 > q^{d-1}.\]
Hence $\dist_\agP\in(a,2m^2a)$ for all agents $\agP$ and the parameters $(m,q,d,\dist)$ are suitable, so that by Theorem \ref{TheoMain}, the {\protname} protocol is informative and perfectly safe for these parameters.
\endproof

As an application, let us return to the example of Section \ref{SecExample}. There were three agents, so we chose $q=4$ and $d=1$. The disadvantage was that the distribution type was noticeably unbalanced, since Alice held the vast majority of the cards. However, as the construction in the proof of Corollary \ref{CorMain} shows, we can actually take $q=3$ provided $d>1$. For $d=2$ and $q=3$, we obtain the distribution type $(18,4,5)$. Observe that Alice holds about four times as many cards as any other agent. In Figure \ref{FigOnlyOne}, we see how this is also true for larger values of $d$. We also see how Alice must hold an increasingly larger portion of the cards as the number of agents rises, but for a fixed $m$, the number of cards she holds grows linearly with respect to the others'.

\begin{figure}
\begin{center}
$
\begin{array}{|c|c|c|c|}
\hline 
\dist & m & q & d\\ 
\hline
\hline 
(18,4,5) & 2 & 3 & 2 \\ 
\hline 
(54,13,14) & 2 & 3 & 4\\ 
\hline 
(162,40,41) & 2 & 3 & 5 \\
\hline 
(486,121,122) & 2 & 3 & 6\\
\hline
\end{array}
\hskip 30pt
\begin{array}{|c|c|c|c|}
\hline 
\dist & m & q & d  \\ 
\hline
\hline 
(48,5,5,6) & 3 & 4 & 2\\ 
\hline 
(192,21,21,22) & 3 & 4 & 3\\ 
\hline 
(100,6,6,6,7) & 4 & 5 & 2\\
\hline 
(500,31,31,31,32) & 4 & 5 & 3\\
\hline
\end{array}
$
\end{center}
\caption{Some choices of suitable parameters. Note that the number of agents is $m+1$ as Alice is counted separately.}\label{FigOnlyOne}
\end{figure}

\section{Concluding remarks}

We have presented a protocol whereby a number of agents holding information that has been privately dealt to them may share it securely even if their communications are intercepted. For convenience of exposition this information is modelled as a deck of cards, but the `cards' may represent any type of sensitive information, such as characters in a password. Our protocol may be used for secret-sharing or other applications that require unconditionally secure aggregation of information, and provides a higher level of security than that in previous work \cite{sadi}.

For future work it may be of interest to consider possible variations or generalizations, for example based on a wider class of combinatorial designs. There are several advantages that such variations might have. First of all, our protocol requires for one agent to hold a large portion of the deck, so it would be convenient to find solutions that work for a larger class of distribution types. Second, we may be interested in obtaining an even higher level of security; \cite{swanson:2012} considered the notion of {\em $k$-perfect security,} where the probability that a given agent holds a set of at most $k$ cards does not change after the agents' announcements. In the two-agent case this is stronger than perfect safety (i.e., $1$-perfect security) when $k>1$, and a multi-agent generalization might also be fruitful. Finally, we mention that solutions which allow Eve to hold cards would be of interest, as finding protocols for such a setting could be useful for applications where portions of the private information has been compromised by the eavedsropper.

%% not want to use model1b-num-names.bst.





% \end{thebibliography}
\end{document}